# Transition between cooperative emission regimes in giant perovskite nanocrystals

*Etsuki Kobiyama[1], Gabriele Rainò[2,3,*], Yuliia Berezovska[2,3], Chenglian Zhu[2,3], Simon C. Boehme[2,3], Maryna I. Bodnarchuk[2,3], Rainer F. Mahrt[1], Maksym V. Kovalenko[2,3,*], Thilo Stöferle[1,*]*

[1] IBM Research Europe – Zurich, Rüschlikon, Switzerland

[2] Institute of Inorganic Chemistry, Department of Chemistry and Applied Bioscience, ETH Zurich, Zurich, Switzerland

[3] Laboratory of Thin Films and Photovoltaics, Empa — Swiss Federal Laboratories for Materials Science and Technology, Dübendorf, Switzerland

* Corresponding Authors: rainog@ethz.ch, mvkovalenko@ethz.ch, tof@zurich.ibm.com

**Abstract**

Interactions between emitters within an ensemble can give rise to cooperative processes that significantly alter the properties of the emitted light. One such process is superfluorescence (SF), where excited electric dipoles spontaneously couple coherently and effectively radiate as one macroscopic emitter. It requires low energetic disorder, high temporal coherence and oscillator strength, and sub-wavelength volumes of material can be sufficient. Conversely, amplified spontaneous emission (ASE) originates from an avalanche-like stimulated amplification of initially spontaneously emitted photons and does not necessitate temporally coherent interactions among the emitters, but rather requires spatially long enough light propagation within the material to harvest the optical gain. Cesium lead halide perovskite nanocrystals (NCs) are one of the very few materials where both ASE (in disordered films) and SF (in ordered assemblies) were observed, however leaving unclear whether and how these regimes could be connected. Here, we demonstrate that temperature and excitation density can drive the transition between both regimes in a thin film of giant $CsPbBr_3$ perovskite NCs. At temperatures below 45 K, excitonic SF was observed, whereas above a transition range between 45 K and 100 K, ASE prevails, but requires increased optical excitation and emitter density. Our results work out the different collective effects present in lead halide perovskites, providing fundamental insights into cooperative phenomena and important guidance for the development of compact and bright (quantum) light sources.

**Introduction**

Lighting and display applications are almost exclusively making use of photoluminescence (PL), where large numbers of emitters radiate light essentially as independent entities (**Fig. 1**a). Cooperative effects, however, would allow a much higher photon flux and spectrally more brilliant emission. Most prominently, laser devices[1] deliver coherent emission by placing the optically active material within a resonator where the light is amplified in an avalanche-like process while circulating in it. However, the necessity to have a low-loss optical resonator, which is tuned to the material absorption/emission wavelength, makes lasers comparably complicated, expensive, and bulky. The stimulated emission process that is harnessed in lasers can also be used without resonator to build sources where the light output is spectrally much broader and less coherent. In such a waveguiding configuration, which typically is even less compact than a laser, a few initial, spontaneously emitted photons become amplified through stimulated emission during their propagation within the excited material, which must provide high optical gain and low internal losses during propagation. In this amplified spontaneous emission (ASE) process, the photon flux increases exponentially before saturation is reached for larger propagation length. The effective coupling between the emitters in ASE, or stimulated emission in general, is induced by propagating photons, while the quantum mechanical phases of the emitters do not need to be coherent (**Fig. 1**b).

Alternatively, superfluorescence[2,3] (SF) can occur when the emitters have synchronized quantum mechanical phases, effectively behaving as one giant emitter instead of an uncorrelated ensemble. Even if the emitting dipoles are initially excited incoherently, they can spontaneously develop macroscopic coherence through coherent interactions via the common electromagnetic field and collectively emit a light burst of correlated photons (**Fig. 1**c). In contrast to ASE, a sub-wavelength-sized region is theoretically large enough to harness this cooperative effect, but sufficiently low homogeneous and inhomogeneous broadening and high oscillator strength are necessary. SF exhibits several key characteristics: first, SF features an accelerated decay with a radiative lifetime $\tau_R \propto \tau_0/N$, where $\tau_0$ is the single-emitter radiative lifetime and $N$ is the number of excited coupled emitters; second, the SF pulse peak intensity grows as $I_{peak} \propto N^2$; third, due to the time that the emitters need to spontaneously synchronize, SF emission occurs with a finite delay time after excitation that decreases with increasing excitation fluence as $\tau_D \propto (\ln N)/N$; fourth, the so-called Burnham-Chiao ringing in the time domain can arise from Rabi-like periodic energy exchange between the electromagnetic field and the macroscopic polarization in the material[4,5]; fifth, from a quantum-optics viewpoint, SF pulses exhibit photon bunching, and a controlled number of excited emitters could be used to realize strongly correlated multi-photon states[6].

ASE and SF are based on different physical mechanisms, i.e., stimulated emission and coherent dipole-dipole coupling between excited emitters, respectively, and transitions between these regimes may occur[7]. The phase diagram sketched in **Fig. 1**d illustrates the dominant emission regimes depending on excitation density and temperature. SF emission can be observed when the boundary condition for the dephasing time $T_2 > \sqrt{\tau_D \tau_R}$ is satisfied. For materials where $T_2$ at low temperatures is on the order of the single-emitter radiative decay time $\tau_0$, the condition is already met for two coupled emitters. With increasing temperature, phonon-induced dephasing reduces $T_2$ and therefore involves increasingly higher excitation strength to maintain SF, consistent with the $N$-dependencies of $\tau_R$ and $\tau_D$ Once the SF requirement of slow decoherence cannot be fulfilled any more, the emission occurs incoherently, as either ASE or standard PL. The threshold from PL to ASE is given when the optical gain accumulated by propagation through the excited material within the waveguide exceeds the optical losses from scattering and absorption. In the time domain, the transition from PL to ASE is characterized by an abrupt change in the emission lifetime with increasing excitation density, as described by the different Einstein coefficients for the spontaneous and stimulated emission rates, with the latter depending additionally on the photon density. At a given pump fluence, decreasing the density of emitters reduces both the number and density of coherently coupled excitons for the SF regime and the modal gain in the ASE, therefore shifting the boundaries, as described by the dashed lines in the phase diagram **Fig. 1**d.

Whereas ASE does not require temporal coherence among emitters and has been demonstrated in many material classes, SF has only been observed in a few, selected systems such as hydrogen fluoride gases[8], $O_2^-$ centers in KCl crystals[9], CuCl quantum dots (QDs)[10], InGaAs/GaAs multiple quantum wells[11,12], and NV centers in diamond[13]. However, the transition between the two regimes has only been observed in a singular system[7], and only via temperature-induced dephasing. Recently, lead halide perovskites have attracted enormous attention due to their exceptional optical properties[14–19] and technological relevance[20,21]. Their low inhomogeneous and homogeneous broadening as well as their high oscillator strength enabled the observation of SF signatures in ordered perovskite nanocrystal (NC) superlattices[22–24], in some instances also in films even up to elevated temperatures[25,26]. In different configurations, efficient ASE has been realized in thin films[27–29], but it has remained an open question whether the very same perovskite material system can support both cooperative regimes and how transitions between such regimes may be experimentally induced and observed.

Here, we report transitions between emission regimes of SF, ASE, and standard PL in thin films of giant $CsPbBr_3$ perovskite NCs. Phase boundaries have been identified by systematically altering the exciton dephasing time by the sample temperature, the number density of excitons by the employed excitation fluence, the density of NCs by the dilution of NC solutions during film preparation, and the photon propagation distance through the pumped stripe length. These results provide comprehensive insights into the different emission mechanisms and transitions between them, which are important for the fundamental understanding of cooperative emission processes and the development of compact, ultra-bright light sources.

**SF in ensembles of giant NCs at 6 K**

We synthesized $CsPbBr_3$ NCs via the ligand-assisted reprecipitation (LARP) technique and prepared thin-film samples via drop-casting on $Si/SiO_2$ substrates, see Methods. A transmission electron microscopy image shows a typical particle size of 40 nm × 40 nm × 200 nm (**Fig. 2**a), much larger than the exciton Bohr diameter of ~7 nm[20]. These "giant" NCs support excitons in the weak confinement regime with characteristic fine structure and multi-exciton states with binding energies of ca. 15 meV[30]. For our experimental conditions, we calculate that an excitation fluence of 1 µJ cm$^{-2}$ leads to the creation of ~44 excitons per NC, which is a fraction of the Mott density and would correspond to 0.15 excitons per NC for 10-nm-sized NCs. In our measurements at cryogenic temperatures, we excite with fluences in the range 0.2–14 µJ cm$^{-2}$ and increase up to 640 µJ cm$^{-2}$ at room temperature, which is then significantly above the Mott density[31]. The individual giant NCs are within the Dicke regime[2] because the emitting volume $V < \lambda^3$ where $\lambda$ is the emission wavelength. But here we are investigating ensembles of them, so we cannot conclude if the observations involve inter-NC processes. At room temperature, the ensemble shows PL with a single Gaussian emission peak centered at a wavelength of 526 nm and FWHM of 18.1 nm (81 meV), at cryogenic temperature ($T$ = 6 K) the emission red-shifts and narrows down to a Lorentzian shape, the typical spectral shape of a (effective) single dipole, centered at 541 nm with FWHM of 2.8 nm (12 meV) (**Fig. 2**b and Extended Data Fig. 1a). The emission linewidth at cryogenic temperature is significantly narrower compared to those of ensembles of NC in the intermediate confinement regime due to the lower inhomogeneous broadening of the exciton energies, with the homogeneous broadening of individual NCs being <1-2 meV[30]. Below about 50 K, the QY exceeds 70%, and hence, the luminescence decay in this temperature range is predominantly radiative (Extended Data Fig. 1b).

First, we study low-density thin films of giant NCs obtained by dissolving the NCs in a polymer solution prior to dropcasting, see Methods. The typical photoemission dynamics of NC ensembles at 6 K is displayed in Fig. 2c. The decay time decreases gradually with increasing excitation fluence, with the 1/e emission lifetime (top panel of Fig. 2d) exhibiting an up to 10-fold acceleration with respect to the basic emitter decay time of $\tau_0 = 100$ ps, which was obtained from time-resolved PL measurements under a low excitation fluence (see Supplementary Information, Fig. S1a). The almost linear dependence of the time-integrated emission intensity on the excitation fluence (see Supplementary Information, Fig. S2 and S3) shows that this acceleration is not due to quenching by nonradiative decay. The middle panel of Fig. 2d displays the excitation fluence dependence of the peak intensity, $I_{peak}$, best fitted by a nonlinear power-law dependence with an exponent of $\alpha = 1.4$. Although in the Dicke regime the power-law dependence for SF is theoretically quadratic, extended halide perovskite systems have been observed with $\alpha = 1.4 \sim 1.7$ probably due to saturation effects and photon emission from non-SF domains[22,24,26] as well as coupling of emitters beyond the Dicke regime ($V > \lambda^3$) The bottom panel of Fig. 2d shows the shortening of the build-up time $\tau_D$ as a function of excitation fluence. The above discussed excitation-fluence-dependent emission dynamics are typical signatures of SF, which occurs in these giant CsPbBr$_3$ NCs upon 10 times less excitation fluence compared to formerly reported superlattices of 10 nm NCs[22–24,32]. Presumably, this is due to the higher absorption cross section at 400 nm, which is supposed to scale linearly with the NC volume[33], as well as low (in)homogeneous broadening in the giant NCs.

**SF to PL transition in giant NCs**

Since faster phonon-induced dephasing at increasing temperatures may render it increasingly more difficult to build up a collective SF state, we investigate the anticipated transition from SF to standard PL or ASE[7,34] via spectrally- and temporally-resolved emission spectroscopy. The time-integrated emission spectra (Fig. 3a-c) do not exhibit any sign of an (abrupt) transition between regimes. The emission peak continuously shifts towards higher energy with increasing temperature, consistent with the temperature dependence of the bandgap energy of halide perovskites[35,36] (see Supplementary Information, Fig. S2). At the same time, the spectral width increases because of coupling to thermally excited phonons, with the broadening as an indication of decreasing coherence. The emission intensity (see Supplementary Information, Fig. S3) increases almost linearly with excitation fluence in the range 6 – 150 K, suggesting the absence of nonradiative exciton-density-dependent quenching. However, upon approaching room temperature, the overall emission intensity reduces significantly, which we attribute to increasingly efficient nonradiative decay processes, consistent with the observed reduction of the PL QY to ~5% at room temperature (Extended Data Fig. 1b).

In contrast to the time-integrated emission spectra, the time-resolved emission data (Fig. 3d-f and Supplementary Information, Fig. S4) show a pronounced transition in the emission dynamics with temperature. At low temperatures (< 45 K), the emission is consistent with SF, inferred from the accelerated decay with increasing excitation fluence and the observation of Burnham-Chiao ringing. For temperatures between 45 K and 150 K, acceleration of the decay occurs more abruptly and at increasingly higher fluences. Notably, the approximately constant 1/e decay time from PL at lower fluence suddenly drops to the same SF-accelerated decay time like at low temperature. Quantitative analyses of emission dynamics parameters also indicate the change of the emission regime (**Fig. 3**g and Supplementary Information, Fig. S5). The peak intensities $I_{peak}$ follow a power-law with exponents between 1.4 and 1.8 except for the highest temperature (300 K), and the delay time $\tau_D$ of the SF pulse emission exhibits the expected SF fluence dependence only below 100 K. **Fig. 3**h shows the temperature dependence of the decay time for different fluences. We attribute the observed behavior to the decrease of the dephasing time $T_2$ with temperature, therefore the SF condition time $T_2 > \sqrt{\tau_D \tau_R}$ requires a larger number of excited emitters $N$ to be fulfilled, because $\tau_R \propto \tau_0/N$. In these perovskite materials, $T_2$ is comparable with $\tau_0$[14] at low temperatures, and therefore SF starts already from the lowest excitation fluence, whereas at elevated temperatures, an increasingly higher excitation fluence is required for a transition from PL to SF (insets of **Fig. 3**a,b). The trend for the transition between PL and SF is qualitatively similar to the expectation from the $T_2$ time (shaded region in **Fig. 3**h), but it quantitatively differs as this $T_2$ estimate is obtained for much smaller perovskite NCs[14,15] where the decoherence at higher temperature might be different from the giant NCs in this study. At room temperature (**Fig. 3**c), even at very high excitation fluence (>300 μJ cm$^{-2}$), no SF signatures were observed.

**SF to ASE transition in dense NC ensembles**

Next, we study high-density thin films of colloidal giant NCs, obtained by dropcasting polymer-free solutions of NC, see Methods. The layer thickness and effective refractive index is sufficiently high to allow guiding of light in the NC layer. Time-integrated PL spectra of these dense thin films show that below 45 K the single peak structure persists with increasing excitation fluence, whereas above 45 K, firstly a shoulder and then a clearly defined new peak appears with a certain threshold excitation fluence (Fig. 4a-c and Supplementary Information, Fig. S6). This emerging peak, which is not observed with the low-density samples, is red-shifted by up to 30 – 40 meV, significantly larger than any trion or biexciton binding energy[30]. This is typical for ASE because due to the small Stokes shift between absorption and emission, the peak center of the excitonic emission is reabsorbed and optical gain is highest at wavelengths red-detuned from the PL maximum. The spectrally integrated emission intensity increases linearly with excitation fluence (see Supplementary Information, Fig. S7), except for the highest temperature, where at first a sublinear and then a strongly superlinear growth after a certain threshold with an overall S-shape curve, typical for ASE, is observed.

The temporal dynamics (Fig. 4d-f and Supplementary Information Fig. S8) at 6 K are qualitatively similar to the low-density sample, with a gradual emission acceleration with increasing fluence, consistent with SF. However, the fluence-dependent acceleration becomes rather abrupt at higher temperatures. An indication of ASE at temperatures >45 K is the pronounced threshold and subsequent growth of the peak intensity $I_{\text{peak}}$ with a power-law exponent >2 (see Supplementary Information, Fig. S9). This concurs with the temperature where the red-shifted shoulder in the time-integrated spectra appears. Presumably, the temperature-induced dephasing increasingly impedes the establishment of coherent SF, favoring ASE. At 300 K, streak-camera images (Extended Data Fig. 2) indicate that the short emission pulse in the time traces corresponds to the red-shifted emission peak at 537 nm in the PL spectra.

**Experiments and simulations of ASE with variable excitation stripe lengths**

Interestingly, with a much smaller excitation spot (see Methods), no red-shifted peak was observed, even up to much higher fluences at room temperature (Extended Data Fig. 3). In contrast to SF, which needs only (sub-)wavelength-sized volumes, a substantial pumped gain length is required for ASE. To assess quantitatively the optical gain, we performed measurements employing the variable stripe length method[37] (see Methods). As shown in Fig. 5a-c (insets), the red-shifted emission peak appears with smaller excitation fluence for longer excitation stripe length. Fitting the ASE peak area with $I_{ASE} \propto (e^{gL} - 1)/g$ where $L$ is the excitation stripe length[38], yields a modal gain of $g \sim 51 \pm 2$ cm$^{-1}$ (Extended Data Fig. 4). This is lower compared to values for similar CsPbBr$_3$ materials reported in the literature[27,39], but reasonable because the drop-casted thin film is very rough (Extended Data Fig. 5) and therefore, scattering losses are expected to be high.

In terms of the temporal emission dynamics, fast decay components are more pronounced with larger $L$ (Fig. 5a-c), in line with a higher overall gain. What is unexpected, however, is that some of the emission time traces for larger $L$ and higher excitation fluence exhibit very pronounced ringing, resembling the Burnham-Chiao ringing in the SF regime. This contrasts with the assumption of dephasing/decoherence being too fast at room temperature to allow SF. Instead, ASE should be the dominant process, as concluded before from the spectral and time-resolved measurements with the smaller beam and normal-incidence detection (Fig. 4). Indeed, ringing can also manifest in an ASE process, previously observed in host-guest systems and ascribed to the presence of intermediate states[40]. Furthermore, $CsPbBr_3$ has been reported to exhibit delayed optical gain due to slow carrier cooling at high excitation fluence[31]. To probe whether the latter could indeed give rise to an ASE ringing behavior and thus explain the experimentally observed emission dynamics, we performed numerical simulations using a three-level scheme to account for hot carriers (see Methods) (Fig. 5d-f).

The simulations reproduce the emergence of a short emission peak, which corresponds to ASE, and which keeps getting more pronounced by increasing $L$ or/and $N_{frac}$. Here, $N_{frac}$ is the ratio between excited and total number of emitters, which does not scale linearly with excitation fluence due to nonlinear processes such as absorption saturation. Notably, like in the experiments, ASE (in contrast to SF) could not be observed for small excitation beams on the order of tens of micrometers or for low-density thin films (Extended Data Fig. 6). Moreover, the simulation also replicates the pulse ringing for large $L$ and $N_{frac}$ This can be understood by the hot-carrier relaxation continuously filling the emissive band edge state, until its population exceeds the ASE threshold and is depleted in a burst of ASE. Subsequently, it is replenished by relaxing hot carriers and eventually reaches again the ASE regime, effectively producing a ringing behavior. Hence, photon propagation and its concomitant retardation are playing a role for the observed emission dynamics, but not for short $L$ where no ringing is observed. Besides, the absence of the ringing behavior in normal incident excitation (Fig. 4f), can be attributed to the transverse effects during the light propagation[41].

**Conclusions**

We present characteristics and transitions between the different collective emission regimes in halide perovskites that pivotally depend on density, temperature, and excitation area. At cryogenic temperature, we demonstrated that SF occurs in giant perovskite NCs that are much larger than the exciton Bohr diameter, whereas a transition from typical SF signatures to normal PL (for low-density thin films) or ASE (for high-density thin films) was observed at elevated temperatures. Surprisingly, with spatially large excitation beams, a combination of carrier relaxation and propagation effects can mimic SF-like pulse ringing even at room temperature in the ASE regime, suggesting ways to circumvent seemingly similar experimental signatures between SF and ASE. Our results provide essential understanding and guidelines for the interpretation of the peculiar emission dynamics in perovskites, constituting a unique platform to study cooperative phenomena. Furthermore, these can be exploited for the design of ultra-bright perovskite light emitting devices such as quantum light sources and lasers/amplifiers.

**Methods**

**Synthesis and sample preparation**

$CsPbBr_3$ NCs were synthesized in ligand-assisted reprecipitation (LARP) technique following the procedure from *Tamarat et al.*[30] with slight modifications. The hybrid solution was prepared by mixing solutions of 90 µL of $PbBr_2$ (0.67 M in DMF), 465 µL of CsBr (0.043 M in DMF-DMSO in ratio 1:1) and 300 µL of OGB (1 M in DMF), where DMF, DMSO, and OGB denote dimethyl formamide, dimethyl sulfoxide, and oleylguanidinium bromide, respectively. 75 µL of the hybrid solution was injected into a 4 mL vial filled with 2.5 mL of mesitylene under vigorous stirring to initiate fast nucleation of the CsPbBr$_3$ NCs. The green solution with bright green luminescence could be observed in 15-20 seconds. For the purification of CsPbBr$_3$ NCs, 0.25 mL ethyl acetate was added to the 0.5 mL of the crude solution followed by centrifugation for 4 min at 8 krpm (4250 x g). The supernatant was discarded, and the precipitate was kept. For preparing low-density NC thin films, the precipitate was redispersed in 0.4 mL of 1% solution of Poly(1-vinylpyrrolidone)-graft-(1-triacontene) in toluene and 30 µL of the resulting solution of CsPbBr$_3$ NCs was dropcasted on Si/SiO$_2$ substrates and dried under vacuum. For preparing high-density NC thin films the precipitate was redispersed in 0.4 mL of toluene and 30 µL of the resulting solution of CsPbBr$_3$ NCs was dropcasted on Si/SiO$_2$ substrates and dried under vacuum.

**Optical spectroscopy**

Time-integrated and time-resolved photoluminescence has been measured by mounting the sample in a helium exchange-gas cryostat which operates in a temperature range from 6 K to 300 K. As excitation source we used a frequency-doubled regenerative amplifier running at 400 nm with a repetition rate of 1 kHz, delivering pulses of about 150–250 fs duration. To prevent parasitic excitation light, short-pass filters (cut-off wavelength of 442 nm) were used. For both excitation and detection, we used the same focusing lens with 100-mm focal length, resulting in an excitation spot size of about 200 µm in diameter. The recorded photoluminescence was spectrally filtered by means of a long-pass-filter (cut-off wavelength of 480 nm). For the time-resolved measurements, the emission was dispersed by a 150 lines per mm grating in a 0.3-m-long monochromator and detected with a streak camera with nominal time resolution of 2 ps and instrument response function full-width at half-maximum (FWHM) of 4 ps. The time-integrated photoluminescence spectra were recorded by a 0.5-m-long spectrograph equipped with a 300 lines per mm grating and a nitrogen-cooled charge-coupled device camera.

For excitation with a much smaller excitation beam, the laser was focused with a 100X objective lens (Mitutoyo) to a Gaussian spot diameter with 3.5 µm FWHM.

For the variable stripe length measurements, the samples were cleaved to allow the light to be collected from the cleaved edge. The stripe-shaped beam spot was obtained by focusing the excitation laser pulse with a cylindrical lens to 40 µm × 2000 µm FWHM, with both dimensions well-approximated by Gaussians, as confirmed by imaging the excitation spot on a camera. The stripe length was controlled by partially blocking the light with a razor blade on a micrometer stage.

**Superfluorescence fit model**

Assuming that the initial number of coherently coupled dipoles is proportional to the excitation fluence $F$ with the proportionality constant $\zeta$, the power-dependent effective decay time was fitted with

$$\tau_{\text{SF}}(F) = \frac{\tau_0}{\zeta \cdot F + 1} + y_0. \tag{1}$$

Here, $\tau_0$ was fixed to the values which were obtained from the time-resolved PL measurements of NCs at weak excitation fluences (see Supplementary Fig. S1). $y_0$ was inserted to account for effects such as the finite time resolution. Regarding the SF build-up, we fitted the power-dependent delay time data with the following function based on the assumption that $\tau_D \propto (\ln N)/N$ (with scaling factor $A$) in the SF regime,

$$\tau_D = y_{\text{offset}} + A \frac{\ln(\zeta \cdot F + 1)}{\zeta \cdot F + 1}. \tag{2}$$

Here, a constant time-offset $y_{offset}$ is included in the fitting function because the absolute time of the arrival of the excitation pulse cannot be measured reliably from the streak camera. Herein, we used a fixed value of ζ, which we obtained from the fit of the 1/e decay lifetimes. To obtain the absolute time delay, we subtracted the obtained fitting parameter values of $y_{offset}$ from the experimental data points.

**Computational simulation of ASE**

As the excitation energy is much larger than the bandgap energy of the NCs, it has been observed that photoexcited carriers may take several picoseconds to relax into the band edge state[31,42]. Therefore, excitation energy well above the bandgap causes an effective delay in the pumping process from the ground state to the band edge state, i.e., the emitting state. Moreover, it is known that the competition between stimulated emission and a delayed pumping process can lead to ringing of ASE emission pulses[40]. We simulated the ASE dynamics under variable stripe length excitation. First, the propagation of electromagnetic waves in media is described by the following Maxwell's equations:

$$\nabla \times E = -\mu_0 \frac{\partial H}{\partial t}, \nabla \times H = \varepsilon_0 \frac{\partial E}{\partial t} + \frac{\partial P}{\partial t}. \qquad (3)$$

Here, *E(t, x)* and *H(t, x)* are the electric and magnetic field with the electric and magnetic constants $\varepsilon_0$ and $\mu_0$, respectively. *P(t, x)* represents the sum of classical oscillator dipoles in the medium. For simplicity, we only considered the propagation of a transverse electromagnetic (TEM) wave in *x* direction.

To replicate the delayed pumping and emission process of the perovskite material, we employed a three-level system as the medium. Carriers are excited from the ground state to the high energy state and then relax to the band edge state non-radiatively. A photon is emitted when a carrier relaxes from the band edge state to the ground state. This system is described with the following pair of rate equations[1]:

$$\frac{dN_h}{dt} = -\frac{1}{\tau_h}N_h \tag{4}$$

$$\frac{dN_e}{dt} = \frac{1}{\tau_h}N_h - \frac{1}{\tau_e}N_e + \frac{1}{\hbar\omega}E\cdot\frac{dP}{dt} \tag{5}$$

$$\frac{dN_g}{dt} = \frac{1}{\tau_e}N_e - \frac{1}{\hbar\omega}E\cdot\frac{dP}{dt} \tag{6}$$

$N(t, x)$ is the number density of carriers at the given time $t$ and position $x$. The subscripts h, e, and g correspond to the high energy state, the emitting band edge state, and the ground state. $\tau_h$ is the time constant of the hot carrier relaxation process from the high energy state to the band edge state, and $\tau_e$ is the spontaneous emission lifetime from the band edge state to the ground state. $\hbar\omega$ corresponds to the energy separation between the band edge state and the ground state.

According to the Lorentz model, the interaction between the electric field and the polarization in the media can be described as the following equation:

$$\frac{d^2P}{dt^2} + \Delta\omega\frac{dP}{dt} + \omega^2 P = \kappa\Delta NE. \tag{7}$$

$\Delta\omega$ is the linewidth of the transition energy between the band edge state and the ground state. On the right-hand side of the equation (7), $\kappa = e^2/m$, where $e$ is the elementary charge and $m$ is the reduced mass of the electron. $\Delta N$ is the population difference and defined by $\Delta N(t, x) = N_g(t, x) - N_e(t, x)$

Since ASE is initiated by spontaneous emission, we introduced an artificial source of the electric field to simulate spontaneous emission numerically[43]. The artificial source is described as follows:

$$E_{\text{source}}(t,x) = \sqrt{N_e(t,x)\hbar\omega\beta\eta} \cdot \sin(\omega t + \phi_{\text{random}}) \tag{8}$$

Here, $\eta$ is the wave impedance. $\beta$ is a phenomenological parameter which represents the fraction of photons that is spontaneously emitted in the ASE mode, and $\beta$ is set as 0.01 in our simulation. At each given position $x$, the artificial source has an individual random phase $\phi_{\text{random}}$

Fig. 5d-f show the calculated emission intensity time traces. The total number density of the emitters in the system was set as $N_{\text{total}} = 10^{18}$ cm$^{-3}$, and $N_{\text{frac}}$ indicates the fraction of the excited emitters to $N_{\text{total}}$. The calculation qualitatively reproduced the experimental results. First, when the system reaches the population inversion, the time traces show a fast emission burst because of ASE. The emission burst peak intensity increases with the excitation stripe length because the emitted light is increasingly amplified for longer excitation stripe length. Besides, the ringing behavior of the emission pulse is also reproduced well in the calculation. It shows that pulse ringing in the time domain is not necessary a unique signature of SF but also can occur with ASE due to a combination of relaxation and retardation. Based on the similarities of the experimental and calculation results, we conclude that the observed emission dynamics at 300 K can be explained only with ASE, also because for small excitation spots no acceleration or ringing can be observed at all. ASE behavior is expected when dephasing is much faster than SF time scales, as is anticipated at room temperature.

The set of rate equations was solved numerically by Euler's method. The excitation from the ground state to the high energy state was modeled as an impulsive response, and the spatial profile was homogenous all over the stripe-shaped excitation beam spot. The temporal and spatial calculation step sizes were set as $\Delta t = 1 \times 10^{-16}$ s, and $\Delta x = 3 \times 10^{-8}$ m. The parameters were set as $\tau_e = 6$ ns and $\tau_h = 5$ ps. The value of $\tau_e$ was obtained from emission lifetime measurements at 300 K under weak excitation fluence (Fig. S1h). As for $\tau_h$, we took a value from typical timescale of the hot carrier relaxation in $CsPbBr_3$[31]. Even though it is known that hot carrier relaxation times in lead halide perovskites depend on excitation fluence, we use a constant value in our calculation for simplicity.

**Figures**

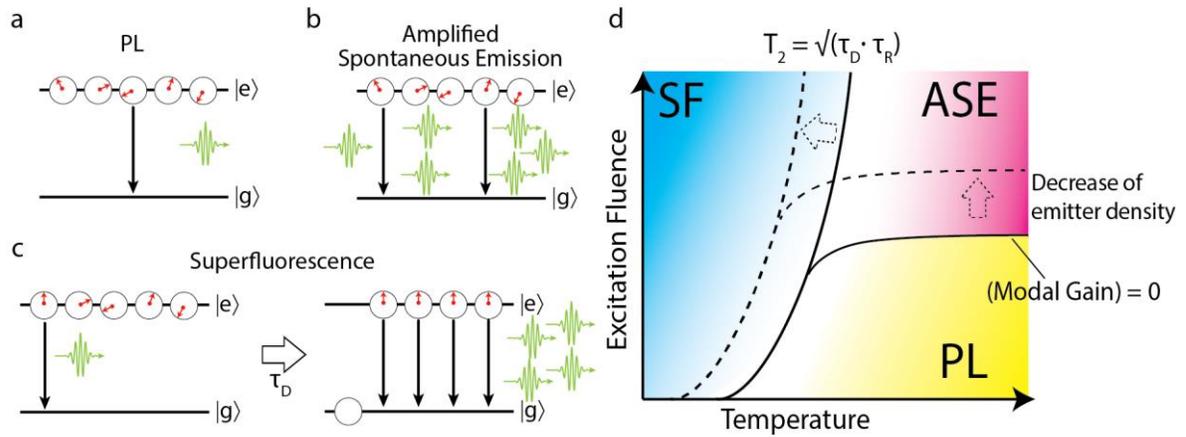

**Fig. 1 | Photon emission processes and phase diagram. a-c**, Illustrations of standard PL (spontaneous emission) (a), amplified spontaneous emission (b), and superfluorescence (c). The excited state and the ground state are described as |e⟩ and |g⟩, respectively. A green waved arrow represents a photon. The white circles represent the emitters with the quantum mechanical phase of excited electric dipoles as red arrows. $\tau_D$ indicates the time that the emitters need to spontaneously synchronize. **d,** Schematic phase diagram of the emission regimes. The solid curves correspond to the borders between different emission regimes, and the dashed curves indicate their qualitative shift when the emitter density is decreased.

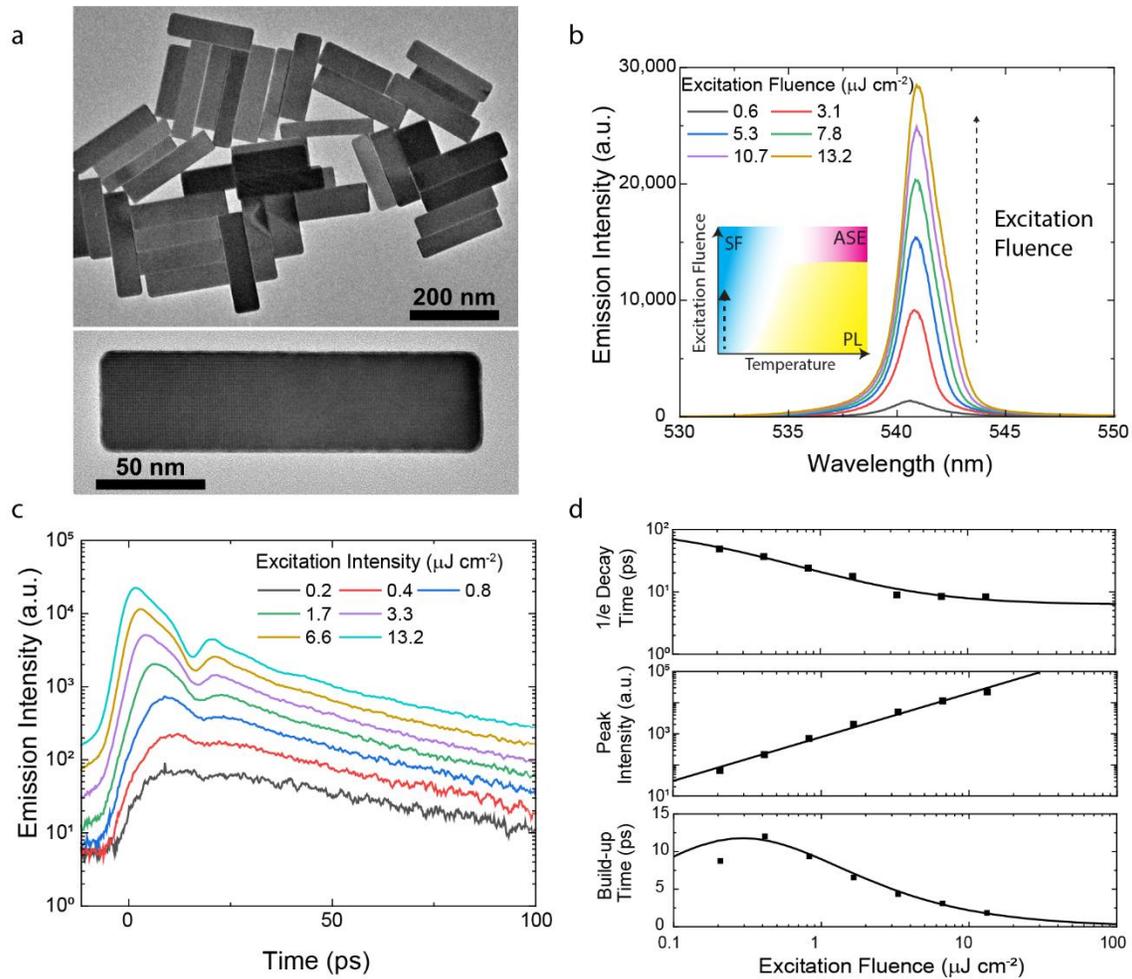

**Fig. 2 | Emission dynamics of the low-density thin film of giant CsPbBr₃ NCs at 6 K. a,** Low-resolution (top) and high-resolution (bottom) transmission electron microscopy (TEM) images of the CsPbBr₃ NCs. **b,** Emission spectra for the several excitation fluences at 6 K. In the inset, the arrow indicates the effective trajectory within the phase diagram with the change of excitation fluence. **c,** Spectrally integrated emission intensity time traces for several excitation fluences. **d,** Extracted data (squares) as a function of the excitation fluence and the SF model curves (solid lines) of: (top) 1/e emission decay time; (middle) time-resolved emission peak intensity; (bottom) emission pulse build-up time.

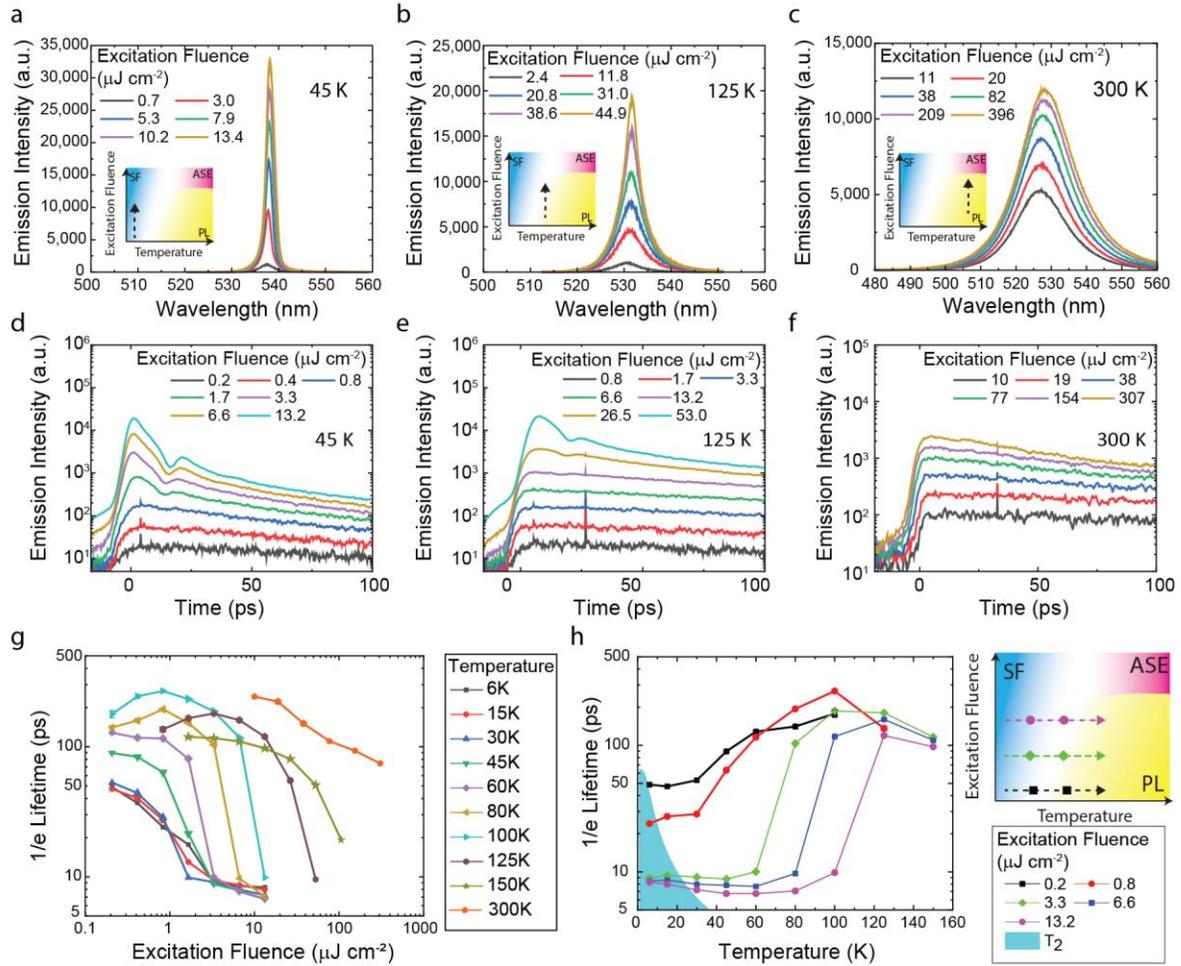

**Fig. 3 | Temperature dependence of emission dynamics of the low-density thin film of giant CsPbBr$_3$ NCs. a-c,** Emission spectra for the several excitation fluences at 45 K (a), 125 K (b), and 300 K (c). The inset indicates the change of the main emission regime in the phase diagram under the change of excitation fluence. **d-f,** Spectrally integrated emission intensity time traces for the several excitation fluences at 45 K (d), 125 K (e), and 300 K (f). **g,** 1/e emission decay time as a function of excitation fluence at different temperatures. **h,** 1/e emission decay time as a function of temperature for the several excitation fluences. The blue shaded area indicates the estimated exciton dephasing time $T_2$ taken from the literature[14,15] for smaller NCs.

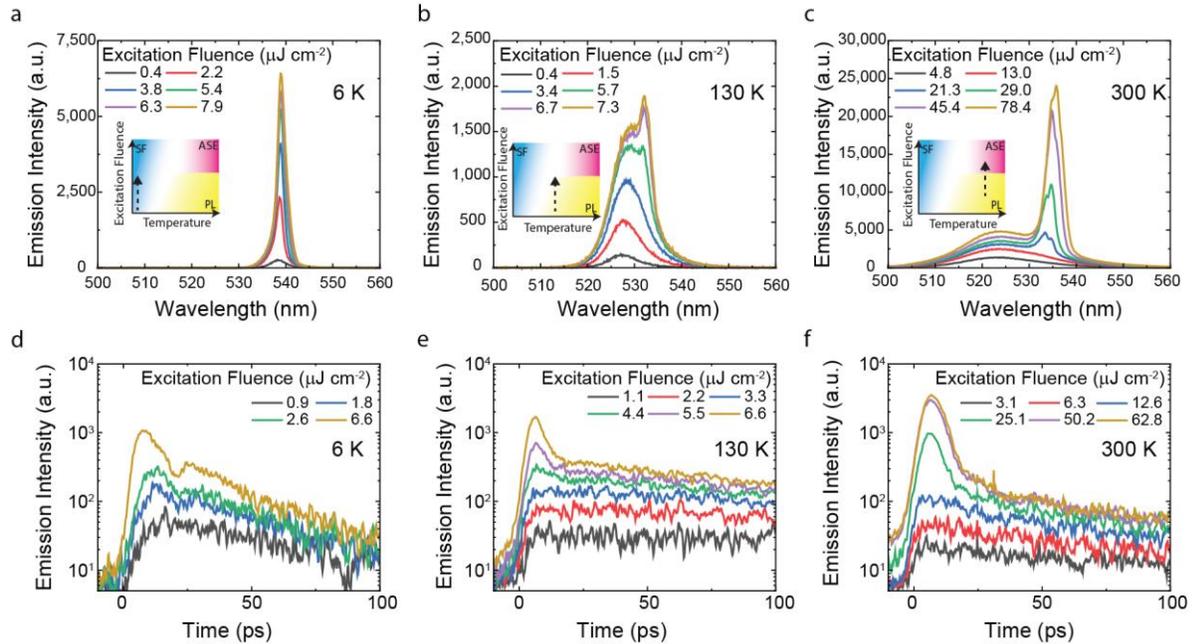

**Fig. 4 | Temperature dependence of emission dynamics of the high-density thin film of giant CsPbBr$_3$ NCs. a-c**, Emission spectra for several excitation fluences at 6 K (a), 130 K (b), and 300 K (c). The inset indicates the sampled emission regimes in the phase diagram with the change of excitation fluence. **d-f**, Spectrally integrated emission intensity time traces for different excitation fluences at 6 K (d), 130 K (e), and 300 K (f).

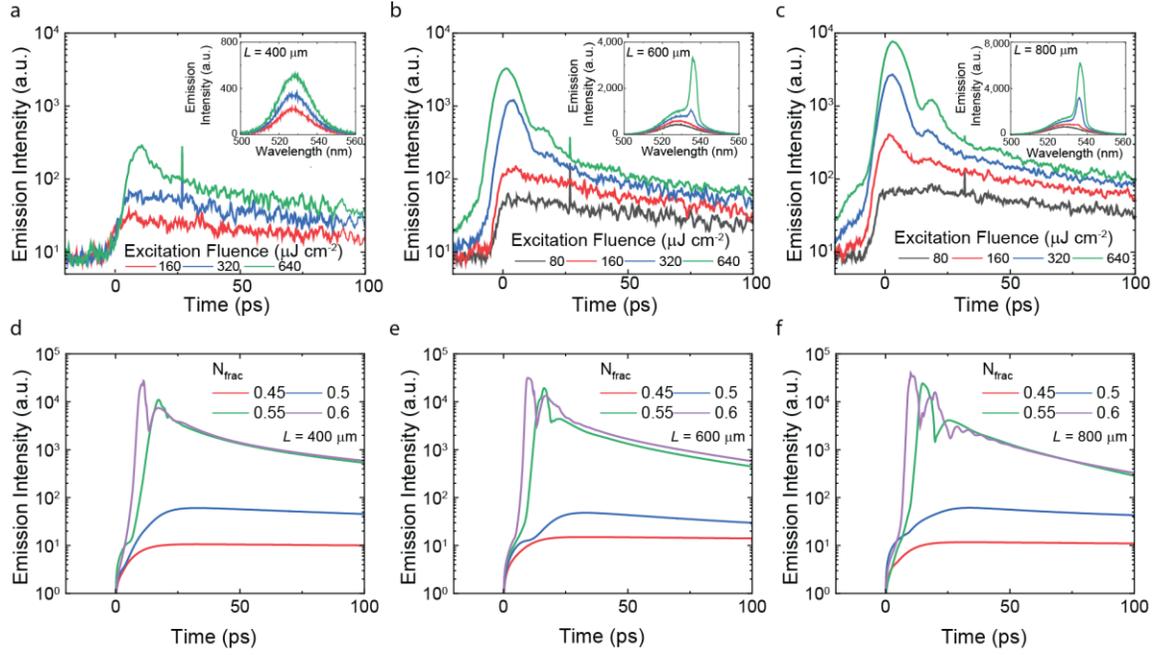

**Fig. 5 | Emission dynamics under variable-stripe-length excitation at 300 K. a-c**, Spectrally integrated time-resolved emission intensity traces detected from the edge of the sample for several excitation fluences with 400 µm (a), 600 µm (b), and 800 µm (c) excitation stripe length $L$. The insets show the corresponding time integrated emission spectra. **d-f,** Numerically simulated emission intensity time traces for different excitation stripe lengths 400 µm (d), 600 µm (e), and 800 µm (f). The total number density of the emitters in the system was set as $N_{total} = 10^{18}$ cm$^{-3}$, and $N_{frac}$ indicates the fraction of the excited emitters to $N_{total}$. The simulation model and parameters are discussed in Methods.

**Extended Data Figures**

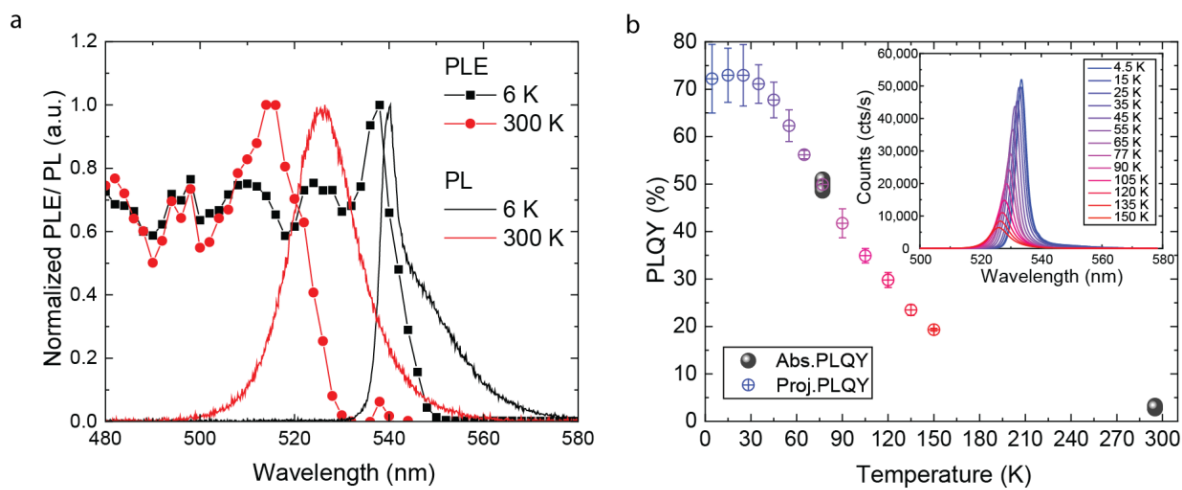

**Extended Data Fig. 1 | Temperature-dependent emission and absorption of giant CsPbBr$_3$ NCs. a**, Normalized PL (solid lines) and PLE (solid lines with markers) of the NCs at 6 K (black) and 300 K (red). **b**, Temperature dependence of the PL QY of a thin film of giant CsPbBr$_3$ NCs. The solid data points show the absolute QY at 77 K and 295 K, obtained via a calibrated absolute-QY spectrometer, and the open circles show the QY at additional temperatures, obtained in a temperature-controlled micro-PL spectrometer and deduced from the relative intensity of the spectrally-integrated PL signal, normalized to the 77 K data point; the error bars represent the uncertainty from temporal and spatial intensity variation across the thin film, and the inset shows the evolution of the emission spectrum with temperature.

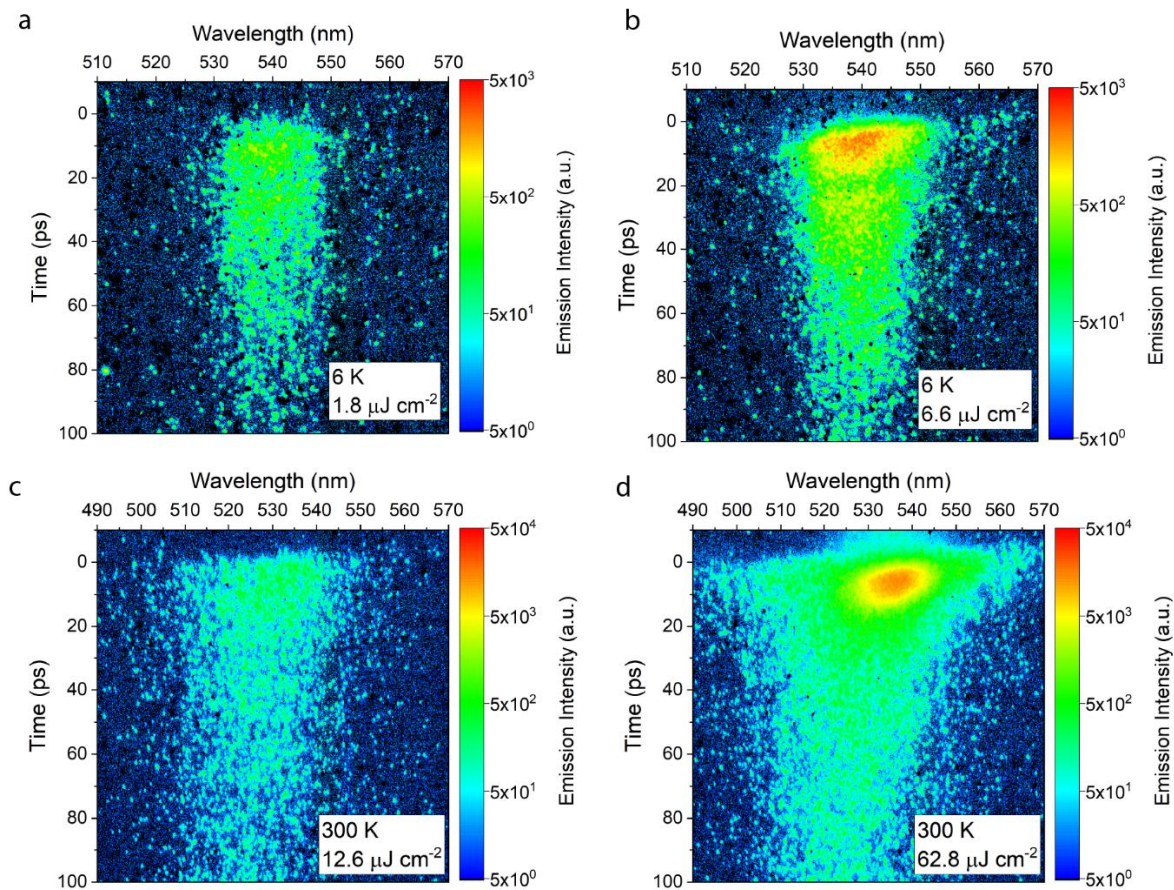

**Extended Data Fig. 2 | Streak camera emission images from the high-density thin film of giant CsPbBr$_3$ NCs.** The sample temperatures are 6 K (a, b) and 300 K (c, d). The excitation fluences are 1.8 µJ cm$^{-2}$ (a), 6.6 µJ cm$^{-2}$ (b), 12.6 µJ cm$^{-2}$ (c) and 62.8 µJ cm$^{-2}$ (d).

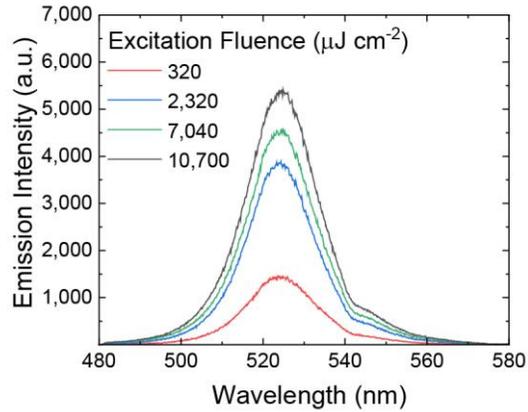

**Extended Data Fig. 3 | Room-temperature emission spectra from the high-density thin film of giant CsPbBr₃ NCs under increasing excitation fluence with a small excitation spot size.** The excitation laser pulse was focused on the sample with a 100x objective lens, and the excitation spot size was 3.5 μm FWHM.

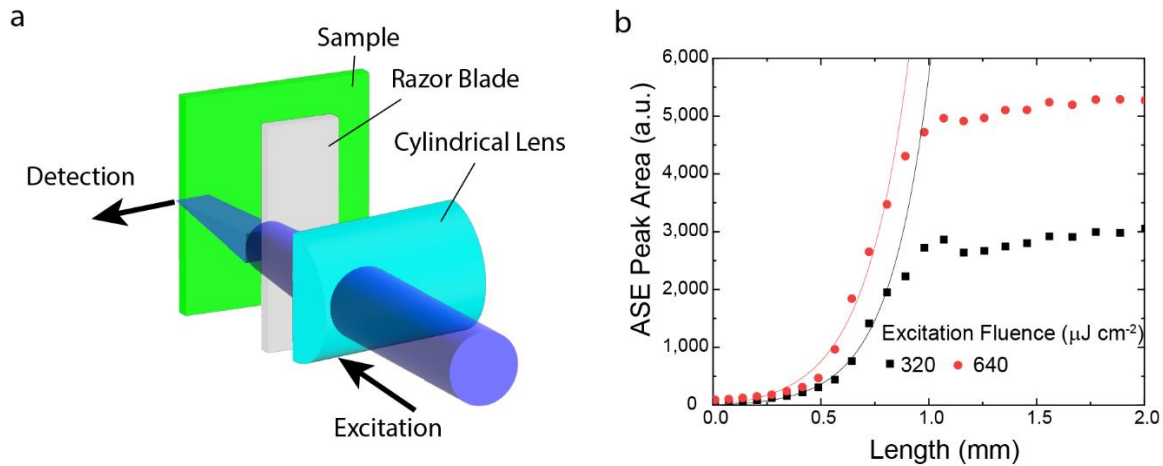

**Extended Data Fig. 4 | Variable-stripe-length excitation measurements. a,** Schematic illustration of the experimental setup. The excitation stripe length was controlled by moving the razor blade. **b,** ASE intensity as a function of excitation stripe length at 300 K for two excitation fluences. The data points show the emission peak intensity collected from the sample's edge. The solid curves are the fitting analysis results with the model function ($I_{ASE} \propto (e^{gL} - 1)/g$), where $L$ is the excitation stripe length and $g$ is modal gain.

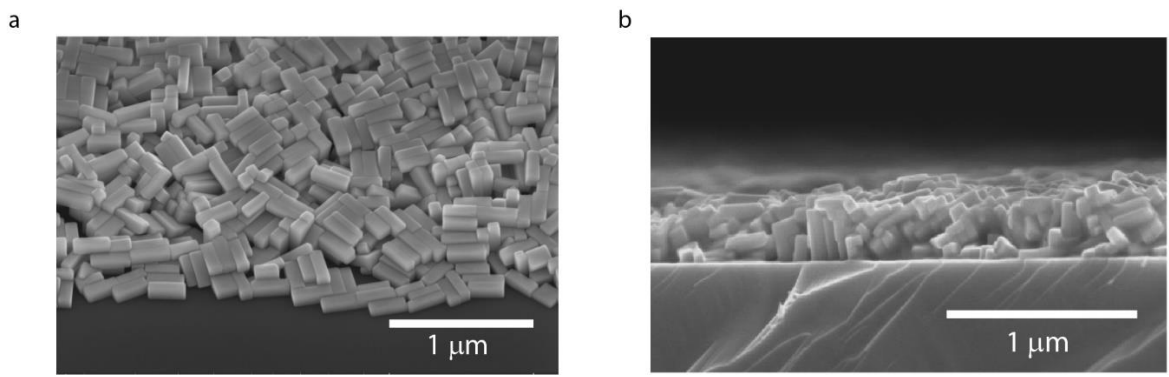

**Extended Data Fig. 5 | Scanning electron microscopy (SEM) image of the high-density thin film. a,** Tilted top-view image of the NC film. **b,** Cross-section image of the NC film from a cleaved edge of the substrate.

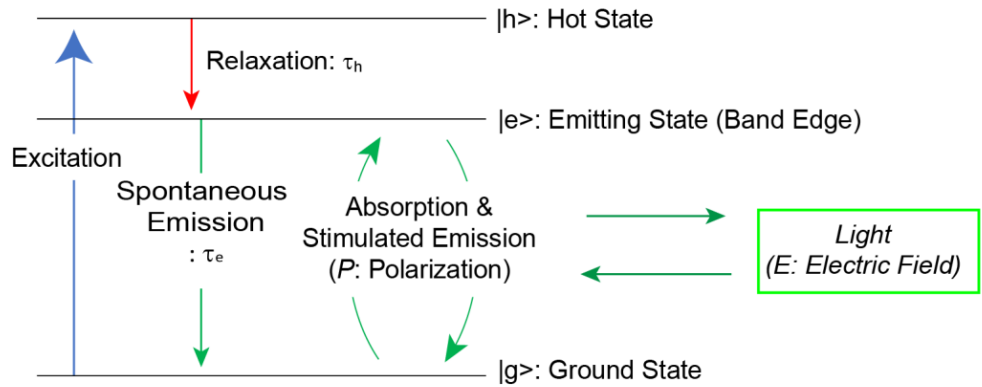

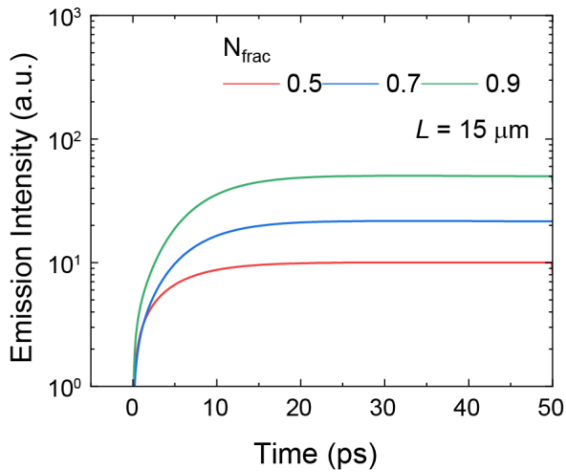
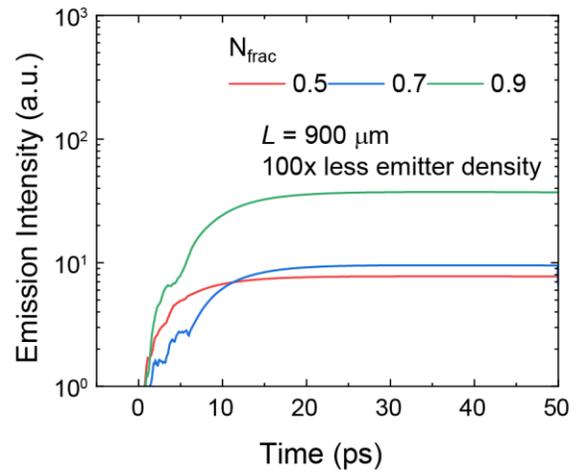

**Extended Data Fig. 6 | Simulated emission dynamics under stripe excitation. a,** Schematic illustration of energy levels and relaxation dynamics of the material system modeled in the simulation. **b,** Numerically simulated emission intensity time traces from the high-density thin film for excitation stripe length of 15 µm, showing no signatures of ASE. **c,** Numerically simulated emission intensity time traces from a low-density thin film for excitation stripe length of 900 µm, showing no signatures of ASE.

**Acknowledgements**

This work was supported by the Swiss National Science Foundation (Grant Number 200021_192308, "Q-Light"), by the Weizmann–ETH Zurich Bridge Program, and by the Air Force Office of Scientific Research and the Office of Naval Research under award number FA8655-21-1-7013.

**Author contributions**

The work originated from continuing interactions between G.R., M.V.K., R.F.M. and T.S. E.K. performed optical measurements, numerical calculations and interpretation of the data with input from G.R., R.F.M and T.S. Y.B. and M.I.B. synthesized giant perovskite NCs and performed the structural characterization. C.Z. and S.C.B performed PL quantum yield measurements. E.K. and T.S. wrote the paper with the contribution from all co-authors. G.R., M.V.K. and T.S. supervised the work.

**Competing interests**

The authors declare no competing interests.

**Materials & Correspondence**

Correspondence and requests for materials should be addressed to Gabriele Rainò, Maksym V. Kovalenko, or Thilo Stöferle.

# Supplementary Information

**Transition between cooperative emission regimes in giant perovskite nanocrystals**

*Etsuki Kobiyama[1], Gabriele Rainò[2,3,*], Yuliia Berezovska[2,3], Chenglian Zhu[2,3], Simon C. Boehme[2,3], Maryna I. Bodnarchuk[2,3], Rainer F. Mahrt[1], Maksym V. Kovalenko[2,3,*], Thilo Stöferle[1,*]*

[1] IBM Research Europe – Zurich, Rüschlikon, Switzerland

[2] Institute of Inorganic Chemistry, Department of Chemistry and Applied Bioscience, ETH Zurich, Zurich, Switzerland

[3] Laboratory of Thin Films and Photovoltaics, Empa — Swiss Federal Laboratories for Materials Science and Technology, Dübendorf, Switzerland

\* Corresponding Authors: rainog@ethz.ch, mvkovalenko@ethz.ch, tof@zurich.ibm.com

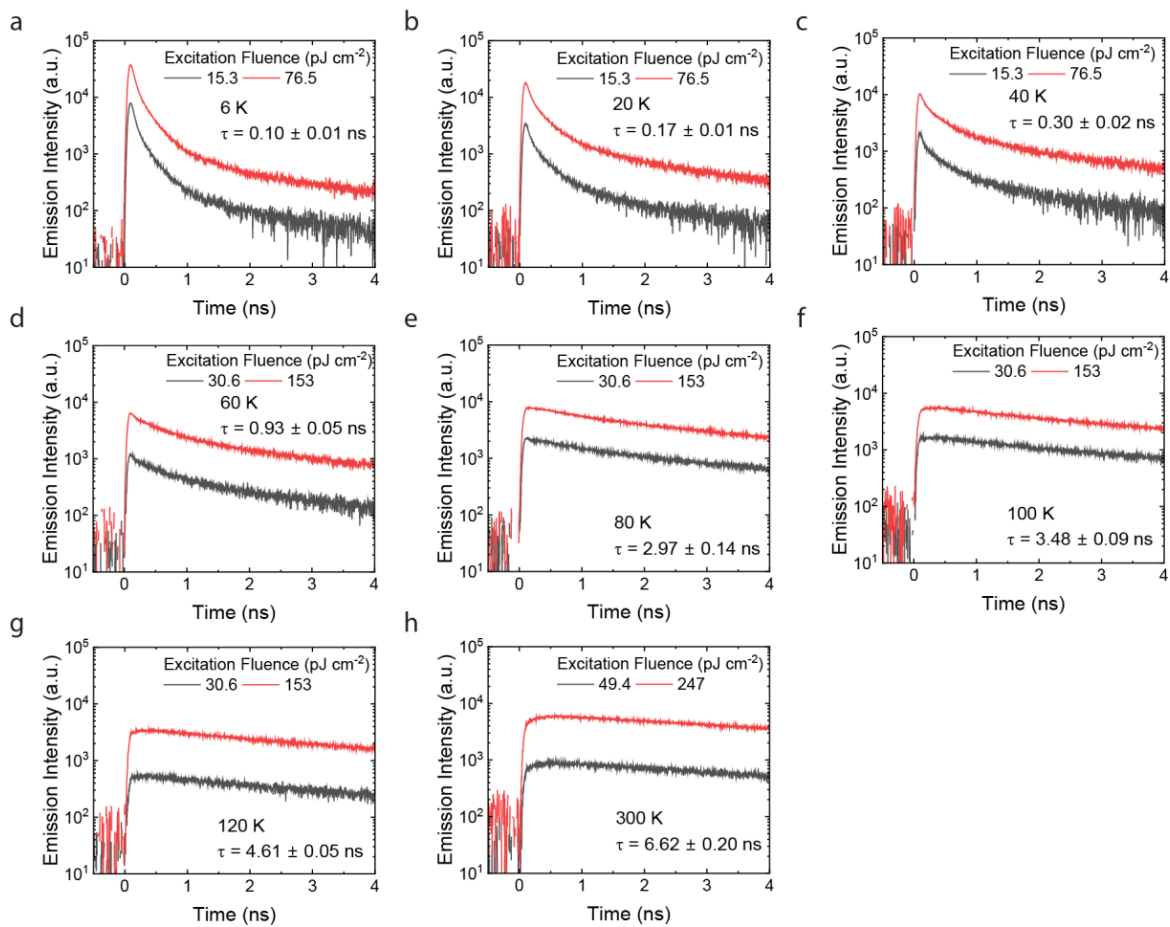

**Fig. S1 | Spectrally integrated PL time traces of the low-density thin film at different temperatures under weak excitation.** The sample temperatures are 6 K (a), 20 K (b), 40 K (c), 60 K (d), 80 K (e), 100 K (f), 120 K (g), and 300 K (h). τ indicates the 1/e emission decay time at each temperature. At 120 K and 300 K the decay time was obtained by curve fitting with a single exponential function since the 1/e decay times are longer than the measurement time window.

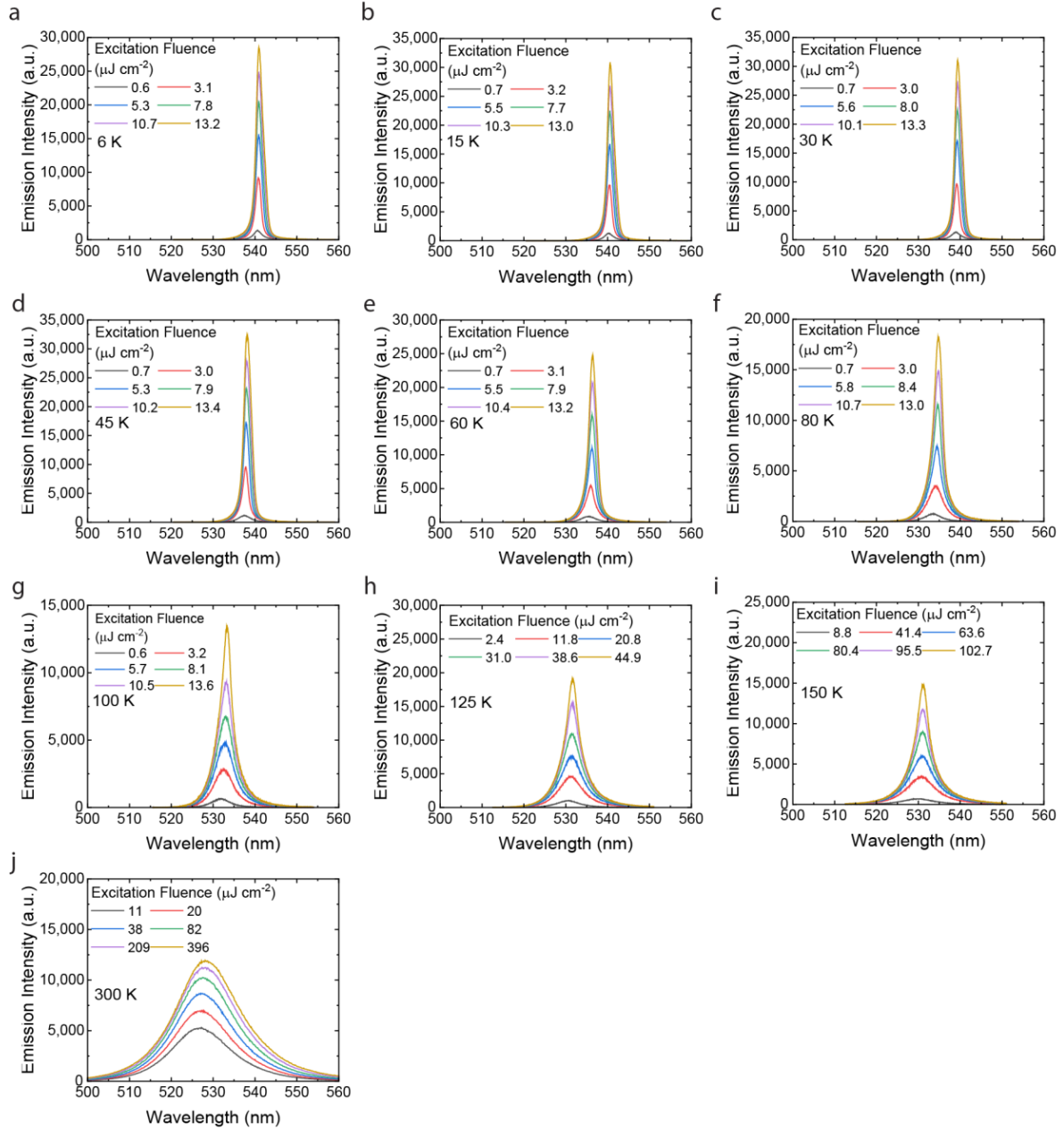

**Fig. S2 | Emission spectra of the low-density thin film under increasing excitation fluence at different temperatures.** The sample temperatures are 6 K (a), 15 K (b), 30 K (c), 45 K (d), 60 K (e), 80 K (f), 100 K (g), 125 K (h), 150 K (i), and 300 K (j).

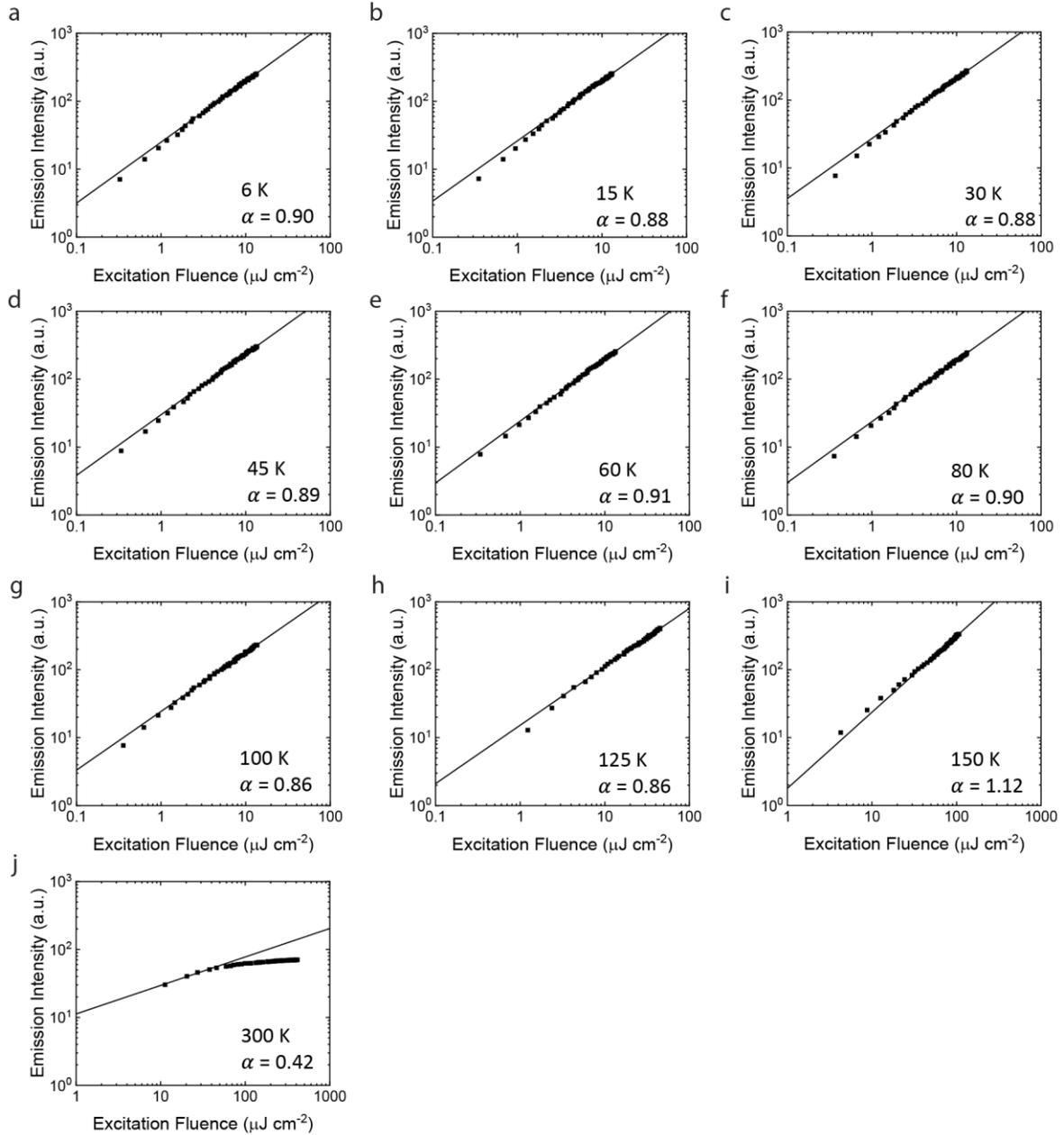

**Fig. S3 | Time-integrated emission intensity as a function of excitation fluence of the low-density thin film at different temperatures.** The sample temperatures are 6 K (a), 15 K (b), 30 K (c), 45 K (d), 60 K (e), 80 K (f), 100 K (g), 125 K (h), 150 K (i), and 300 K (j). α indicates the exponent value obtained from a power-law fit (solid line).

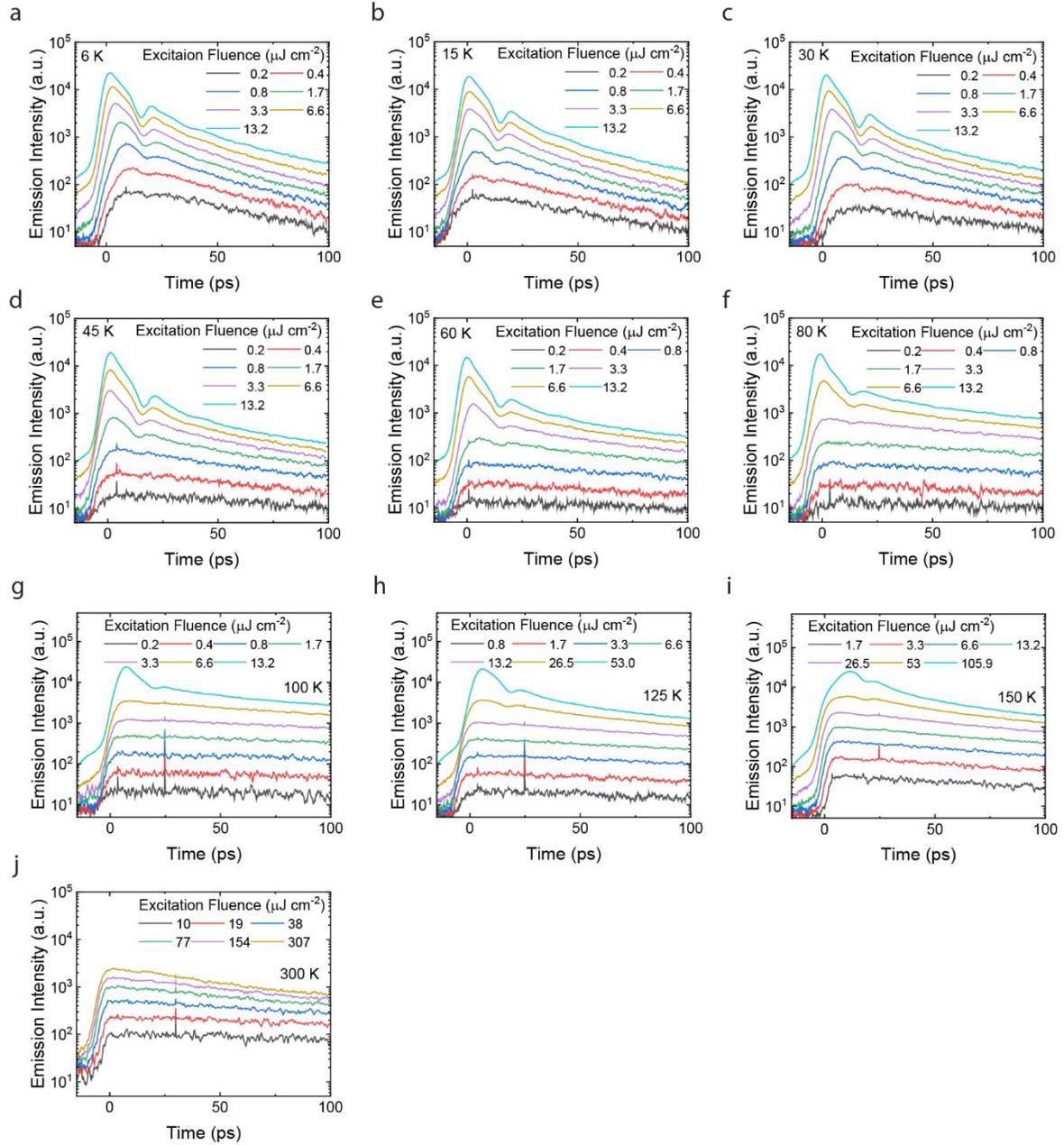

**Fig. S4 | Spectrally integrated emission time traces of the low-density thin film with increasing excitation fluence at different temperatures.** The sample temperatures are 6 K (a), 15 K (b), 30 K (c), 45 K (d), 60 K (e), 80 K (f), 100 K (g), 125 K (h), 150 K (i), and 300 K (j).

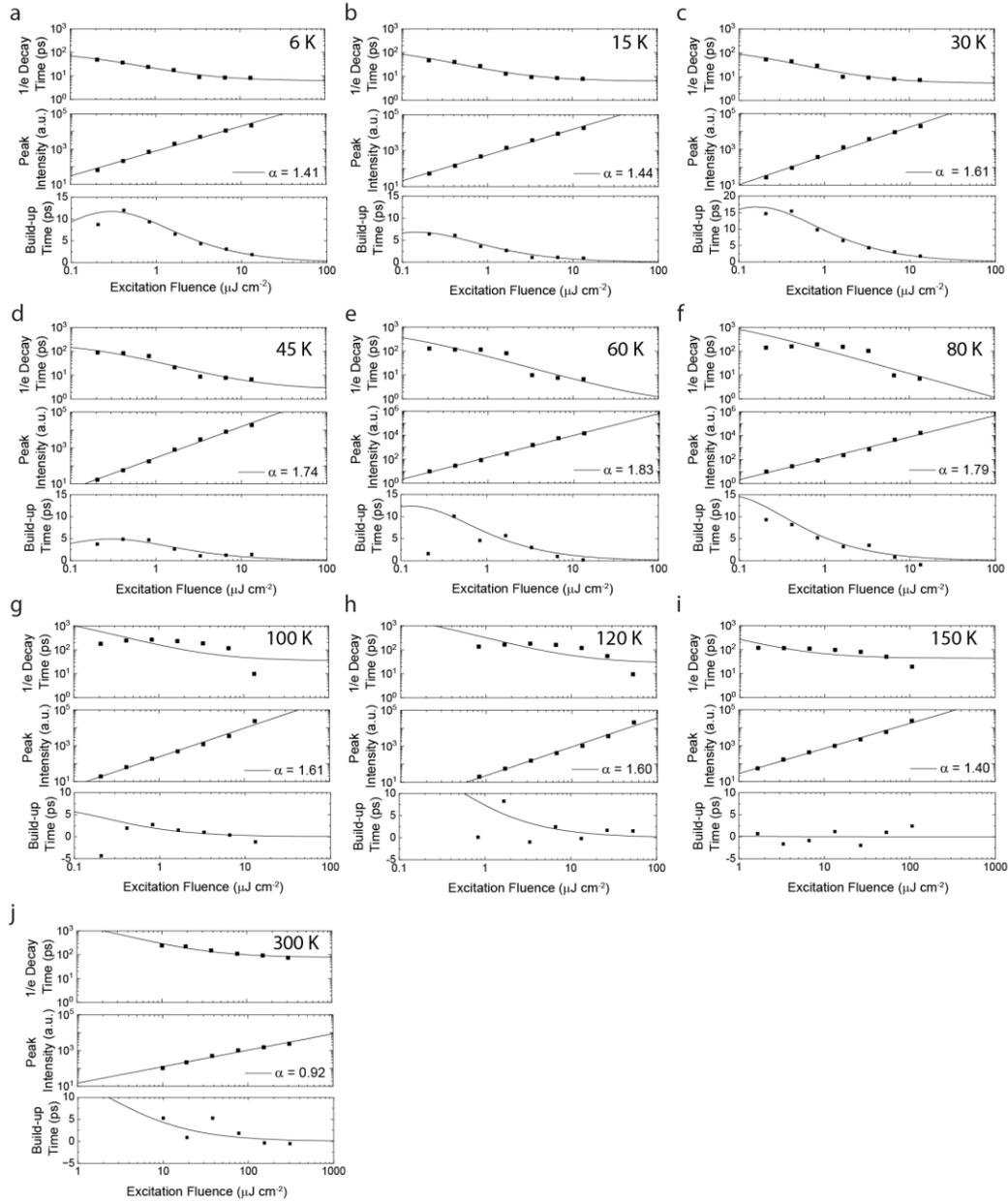

**Fig. S5 | Emission dynamics parameters from the low-density thin film as a function of the excitation fluence and the model curves (solid lines) of: (top) 1/e emission decay time; (middle) time-resolved emission peak intensity; (bottom) emission pulse build-up time.** The sample temperatures are 6 K (a), 15 K (b), 30 K (c), 45 K (d), 60 K (e), 80 K (f), 100 K (g), 125 K (h), 150 K (i), and 300 K (j). The solid lines show best fits with the SF model $\propto \tau_0/N$, where $\tau_0$ is

the individual emitter lifetime as measured at very weak excitation fluence (Extended Data Fig. 2) and *N* is the number of excitations, proportional to the excitation fluence. The model appears to become less and less adequate above 50 K. α indicates the exponent value obtained from a power-law fit (solid line).

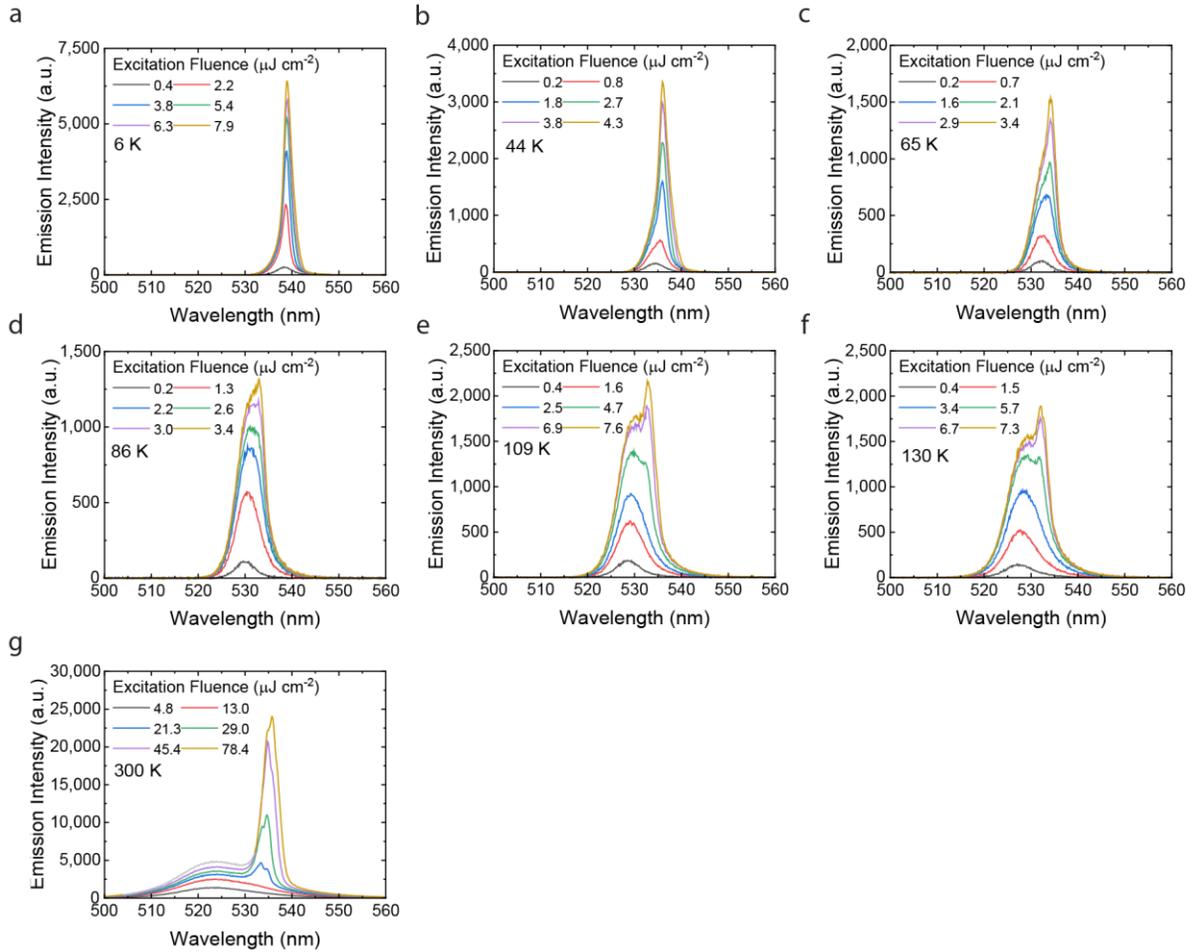

**Fig. S6 | Emission spectra of the high-density thin film under increasing excitation fluence at different temperatures.** The sample temperatures are 6 K (a), 44 K (b), 65 K (c), 86 K (d), 109 K (e), 130 K (f), and 300 K (g).

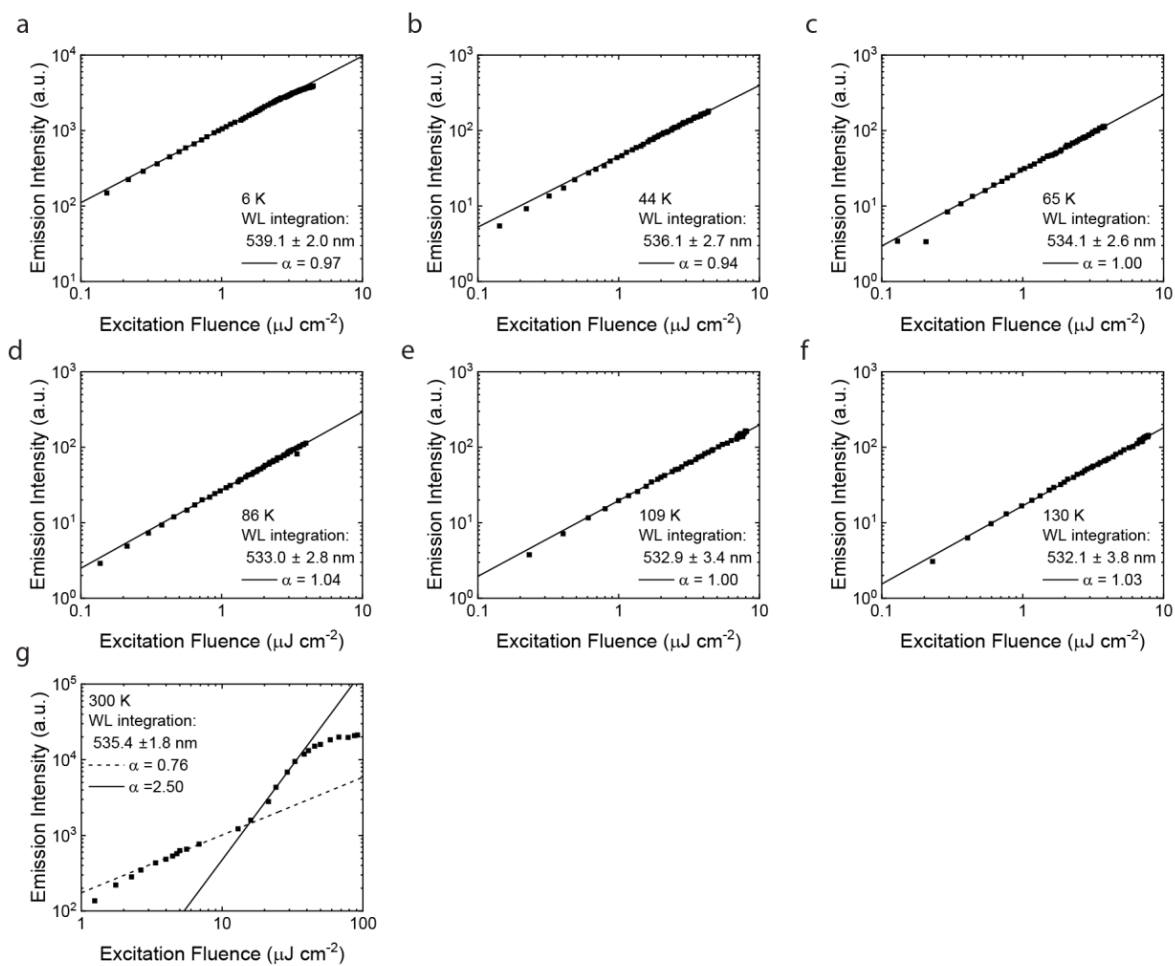

**Fig. S7 | Time-integrated emission intensity of the high-density thin film at different temperatures.** The sample temperatures are 6 K (a), 44 K (b), 65 K (c), 86 K (d), 109 K (e), 130 K (f), and 300 K (g). The spectrum is integrated in a narrow range (specified in each plot) around the single peak (at low temperature), its red shoulder (at medium temperature) or the red peak (at high temperature). α indicates the exponent value obtained from a power-law fit (solid line).

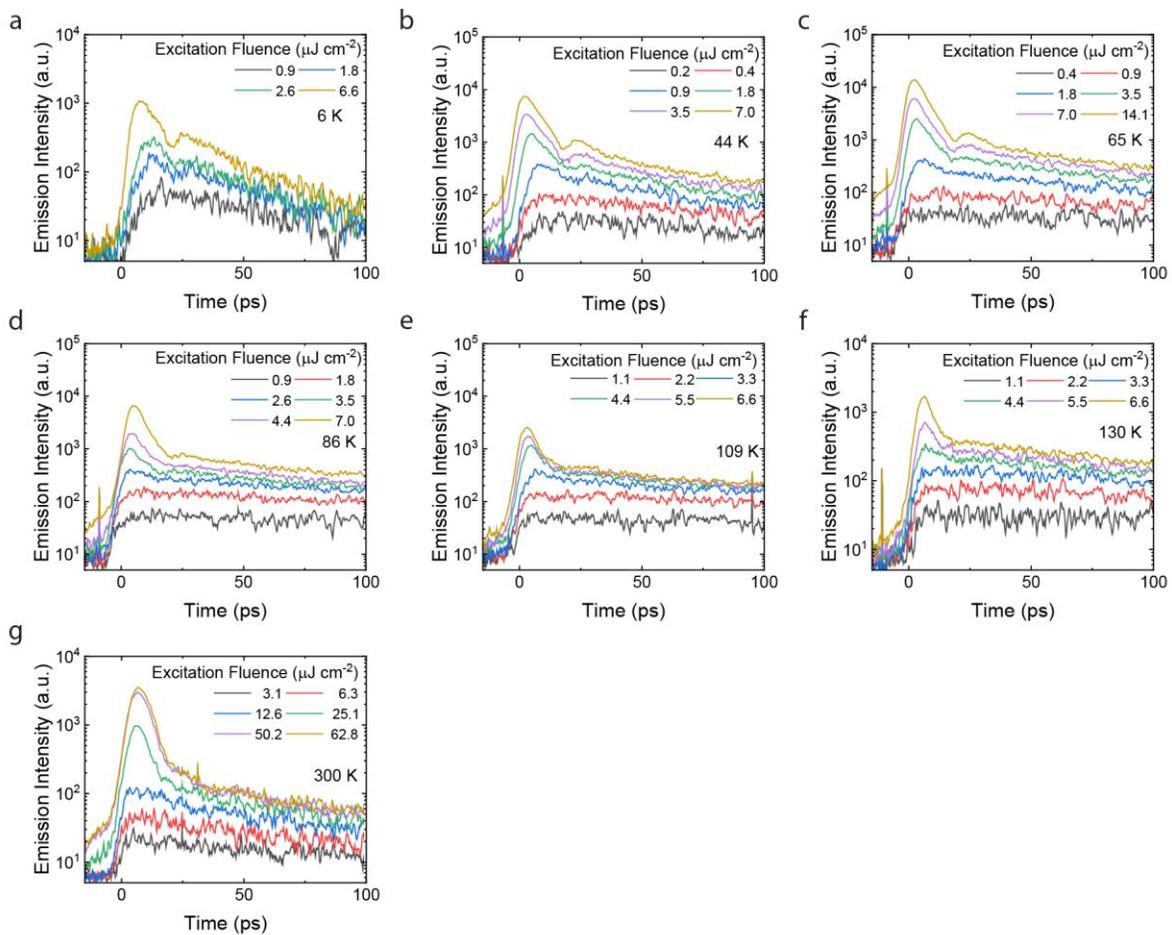

**Fig. S8 | Spectrally integrated emission time traces of the high-density thin film with increasing excitation fluence at different temperatures.** The sample temperatures are 6 K (a), 44 K (b), 65 K (c), 86 K (d), 109 K (e), 130 K (f), and 300 K (g).

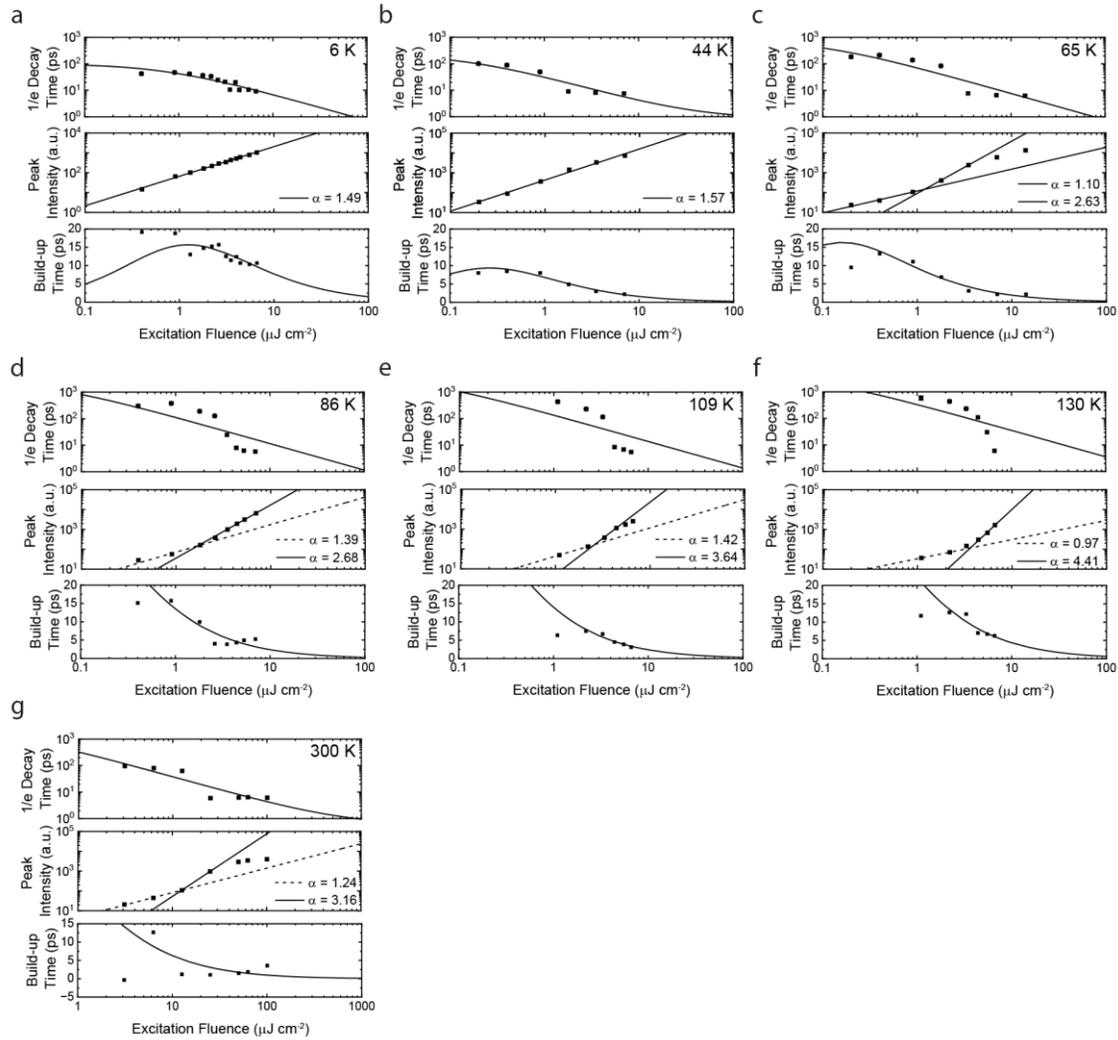

**Fig. S9 | Emission dynamics parameters of the high-density thin film as a function of the excitation fluence and the model curves (solid lines) of: (top) 1/e emission decay time; (middle) time-resolved emission peak intensity; (bottom) emission pulse build-up time.** The sample temperatures are 6 K (a), 44 K (b), 65 K (c), 86 K (d), 109 K (e), 130 K (f), and 300 K (g). The solid lines show best fits with the SF model $\propto \tau_0/N$ where $\tau_0$ is the individual emitter lifetime as measured at very weak excitation fluence (Extended Data Fig. 2) and $N$ is the number of excitations, proportional to the excitation fluence. The model appears to become less and less adequate above 50 K. $\alpha$ indicates the exponent value obtained from a power-law fit (solid line).